\documentstyle[preprint,prl,aps]{revtex}
\begin{document}
\begin{titlepage}
\preprint{\vbox{\hbox{UDHEP-08-97}
\hbox{July 1997}}}
\title{\large\bf NEUTRINO OSCILLATIONS FROM STRINGS AND 
OTHER FUNNY THINGS}
\author{\bf A. Halprin}
\address{Department of Physics and Astronomy, University 
of Delaware\\Newark, DE 19716\\USA}
\maketitle
\bigskip

\begin{abstract}

I will discuss three related unconventional ways to generate
 neutrino oscillations:(1)Equivalence principle violation by 
the string dilaton field (2) Violation of Lorentz Invariance 
and (3)Equivalence principle violation through a non-universal
 tensor neutrino-gravity coupling. These unorthodox oscillation
 mechanisms are shown to be viable at the level of our present 
experimental knowledge and demonstrate that neutrino oscillations 
can probe very profound questions.

\end{abstract}
\bigskip

\leftline{Talk given at the XVI International Workshop on Weak 
Interactions and Neutrinos}
\centerline{Capri, Italy, June 22-28, 1997}

\end{titlepage}

\newpage

In many quarters the unequivocal observation of neutrino oscillations 
is thought to be clear evidence that at least one neutrino has a mass.
  This is by no means the case as will be made clear by the alternate 
and experimentally viable mechanisms to be presented here. As will be 
seen, it is at the present time possible, and perhaps desireable, to 
entertain an entirely different energy dependence for the oscillation 
length than is obtained under the conventional mass mixing hypothesis.

I will discuss three unorthodox ways to interpret the current data on 
neutrino oscillations. Two of them assume neutrinos violate the equivalence 
principle but in very different ways and the third assumes neutrinos 
violate Lorentz invariance, which may instead be viewed as one of the two 
types of equivalence principle violations. The order of the discussion of 
these mechanisms will be inverse to their historical order in the literature.

I begin with recent results obtained in collaboration with C.N. Leung. 
We have examined a scenario proposed by Damour and Polyakov \cite{DP} in 
which the dilaton field of string theory remains massless and does not respect
 the equivalence principle. In their scenario, the  static gravitational potential energy 
between particles A and B a distance r apart is 

\begin{equation}
V(r)=-G_Nm_Am_B(1+\alpha_A\alpha_B)/r
\end{equation}
where $G_N$ is Newton's constant. The term proportional to 
$\alpha_A \alpha_B$ arises from exchange of the spin-0 dilaton 
field, while the other term is the usual universal spin-2 exchange
 contribution. This is much like the Brans-Dicke tensor-scalar
 theory \cite{BD} except that the $\alpha_j$ are not universal, 
thereby violating the equivalence principle.

Leung and I have shown \cite{HL} that even if two neutrinos are
 degenerate in mass (perhaps due to some family symmetry), but 
they interact with gravity in this way, they can still oscillate
 into one another. This is because neutrino A couples to the 
dilaton field with a strength proportional to $\alpha_A$ while
 neutrino B couples with a strength proportional to $\alpha_B$. 
 Consequently, the $\alpha$ basis and the weak basis are not the same, 
and the difference 
can be characterized by some mixing angle $\theta$. We find that
 for two neutrino mixing, say $\nu_e$ and $\nu_{\mu}$, in a constant
 gravitational field the oscillation length is similar to that
 obtained by mass mixing and given by

\begin{equation}
\lambda = \frac{2 \pi E}{\Delta m^{*2}}
\end{equation}
where the effective mass squared difference is  

\begin{equation}
\Delta m^{*2}=-2m_{\nu}^2\alpha_{ext}\Phi_N\Delta\alpha
\end{equation}
$m_\nu$ is the degenerate neutrino mass and $\alpha_{ext}$ is the
 strength with which the dilaton field couples to the matter that
 generates the gravitational field characterized by the Newtonian 
potential, $\Phi_N$. $\Delta\alpha=\alpha_2-\alpha_1$ is the 
difference between the $\alpha$ values of the neutrino species that
 define the $\alpha$ basis. 

For solar neutrinos the limit on $m_{\nu}$ is the experimental 
limit on $m_{\nu_e}$. While there is no severe constraint on
 $\Delta\alpha$, it may be prudent to assume it does not exceed
 the limit on $\alpha_{ext}$, which is 
$\alpha_{ext}^2<10^{-3}$\cite{R}, and constrains the effective
 mass difference to $\Delta m^{*2}<(\Phi_N/10^{-5})10^{-10}(eV)^2$.
 In our region of the solar system the Great Attractor gives the
 largest contribution to the Newtonian potential estimated 
at $3 \times 10^{-5} $\cite{K}. In this case the effective mass
 for solar neutrinos is too small to satisfy the known
experimental limits for an MSW solution but is in the range 
for a vacuum oscillation solution \cite{BK,KP}. With this estimate 
we conclude that in solar neutrino experiments there is no clear 
distinction to be made between this gravitational mechanism and 
conventional mass mixing. If one is less prudent in constraining 
$\Delta\alpha$, then an MSW solution would be possible. In that
 case a distincion between this gravitational mechanism and mass
 mixing might be made by utilizing the 20 $\%$ contribution to 
$\Phi_N$ in the vicinity of the Sun made by the Sun itself,
 which does not contribute at the surface of the Earth.

I now turn to the second oscillation mechanism, a possible 

violation of Lorentz invariance, which has recently been 
introduced by Coleman and Glashow \cite{CG}. They postulate 
that perhaps the limiting velocities of particles are not all 
exactly the same. In that case, $\nu_e$ and $\nu_{\mu}$ might
 each be a superposition of states with limiting velocites 
$v_1$ and $v_2$ characterized by a mixing angle $\theta$. 
Assuming massless neutrinos, this means the energy-momentum
 relation for the $jth$ neutrino is

\begin{equation}
\biggl[ \frac {p}{E} \biggr]_j = v_j
\end{equation}

From this one calculates the  momentum difference for neutrinos 
with a definite energy, E, and obtains the oscillation length 

\begin{equation}
\lambda=\frac {\pi}{E \Delta v}
\end{equation}
where $\Delta v=v_2-v_1$.  Here the oscillation length decreases 
with increasing energy, which is in sharp contrast to 
conventional mass mixing or the gravitational dilaton mixing
 where the oscillation length grows with energy. 

Limits on the mixing parameters dictated by solar neutrino
 experiments have been explored by Glashow 
$\textit{et al}$ \cite{GHKLP} with the result that 
there is an allowed small angle region at the 90 $\%$ confidence
level,

\begin{equation}
\vert \Delta v \vert \sim 6 \times 10^{-19} ,\quad
 0.002<sin^2(2 \theta )<0.003
\end{equation}
and at the same confidence level an allowed large angle 
region,

\begin{equation}
4 \times 10^{-22} \vert \Delta v \vert < 4 \times 10^{-21},
 \quad 0.38<sin^2(2 \theta ) <0.81
\end{equation}
They also find that in contrast to mass mixing, the parameter
 space of the higher energy atmospheric neutrino data overlaps 
that of the solar neutrino data with only two neutrino mixing, 
which makes this mechanism economical in the number of new 
parameters introduced.

Finally I turn to the third mechanism, oscillations induced by
 a violation of the equivalence principle owing to a breakdown
 of universality in the conventional spin-2 gravity-neutrino 
coupling strength. This subject has been discussed in several 
papers during the past few years \cite{HL}. To define the parameters,
 I write the gravity-neutrino interaction (linear approximation)
 for neutrino species $l$ as

\begin{equation}L=f\gamma_lG_{\alpha\beta}T_l^{\alpha\beta}
\end{equation}
where $G_{\alpha\beta}$ is the conventional spin-2 gravitational 
field, $T_l^{\alpha\beta}$ is the energy momentum tensor for the
 $lth$ neutrino species, and $f^2=8\pi G_N$. Universality of this 
interaction is violated by letting $f\rightarrow f\gamma_l$, in 
which the species dependent quantity $\gamma_l$ is close to but 
not equal to unity. For massless neutrinos in a constant Newtonian
 potential, this leads to the energy-momentum relation 

\begin{equation}
p=(1+2\Phi_N\gamma_l)E
\end{equation} 
for the neutrino states that define the $\gamma$ basis. These 
states therefore travel with different speeds, 
$v_l=1+2\Phi_N\gamma_l$. Such a situation is therefore 
phenomenologically identical to the previous case of Lorentz 
invariance violation with $\Delta v=2\Phi_N\gamma_l$, and can 
be viewed as an "explanation" for an apparent violation of
 Lorentz invariance should one become convinced that such a 
violation has occurred. In principle these two mechanisms can
be distinguished by the $\Phi_N$ dependence but it would be
 difficult in light of the dominance by the Great Attractor 
as discussed earlier.

In conclusion, the three oscillation mechanisms presented here
 give lie to the oft repeated statement that the observation 
of neutrino oscillations would be conclusive evidence that at
 least one neutrino is massive and degeneracy excluded. Moreover,
 it  demonstates that the study of neutrino oscillations provides
 a valuable probe for exploring limits on other kinds of fundamental
 questions where nature may still have some surprises left for us.  

\centerline{ACKNOWLEDGEMENTS}

I would like to thank the organizers of this workshop for
 providing such an inspirational atmosphere for the meeting.
This work was supported in part by the U.S. Department of Energy 
under contract DE-FG02-84ER40163.

\end{document}